**Value-based optimization of healthcare resource allocation for COVID-19 hot spots**

Zachary A. Collier[1], Jeffrey M. Keisler[2], Benjamin D. Trump[3], Jeffrey C. Cegan[3], Sarah Wolberg[4], Igor Linkov[3]

[1] Radford University, Radford, VA, USA

[2] University of Massachusetts Boston, Boston, MA, USA

[3] US Army Engineer Research and Development Center, Concord, MA, USA

[4] US Army Corps of Engineers, North Atlantic Division, Concord, MA, USA

**Abstract:** With the emerging COVID-19 crisis, a critical task for public health officials and policy makers is to decide how to prioritize, locate, and allocate scarce resources. To answer these questions, decision makers need to be able to determine the location of the required resources over time based on emerging "hot spot" locations. Hot spots are defined as concentrated areas with sharp increases in COVID-19 cases. Hot spots place stress on existing healthcare resources, resulting in demand for resources potentially exceeding current capacity. This research will describe a value-based resource allocation approach that seeks to coordinate demand, as defined by uncertain epidemiological forecasts, with the value of adding additional resources such as hospital beds. Value is framed as a function of the expected usage of a marginal resource (bed, ventilator, etc.). Subject to certain constraints, allocation decisions are operationalized using a nonlinear programming model, allocating new hospital beds over time and across a number of geographical locations. The results of the research show a need for a value-based approach to assist decision makers at all levels in making the best possible decisions in the current highly uncertain and dynamic COVID environment.

## 1. Introduction

The COVID-19 (SARS-CoV-2) pandemic presents a novel and challenging combination of disease characteristics, including the high risk of inter-human transmission, long incubation times, and presence of asymptomatic carriers, positioning it as a "perfect storm" to impose a strong epidemiological burden on society (Lippi et al., 2020). With the emergence of this pandemic, a central question faced by communities across the globe is how decision makers should prioritize and allocate scarce healthcare resources such as personal protective equipment (PPE) and ventilators (Laventhal et al., 2020; Zaza et al., 2016). Answering the question, "Where will the needs be?" is critical to the successful allocation of resources. From a public health perspective, there is a need to predict the timing and location of areas which will be particularly hard-hit, so that additional resources (e.g., hospital beds, ICU beds, ventilators) can be allocated to lessen the impact of a sudden and rapid influx of admitted patients. Such "hot spots", where a sharp increase in geographically concentrated COVID-19 cases occur, can place a great stress on hospital resources and associated supply chains (Golan et al., 2020).

Hospital beds and ventilators are critical resources for treating patients in any pandemic situation. Excessive demand on existing hospital bed resources results in "capacity strain", which is associated with increased mortality and other adverse health outcomes (Eriksson et al., 2017). The ability to optimize the allocation of scarce healthcare resources in areas which are experiencing, or will likely experience, a surge in demand exceeding capacity, can help to ease the burden on the healthcare system in that area (Meyer et al., 2020).

One resource allocation strategy is to follow a "needs-based" approach, which involves identifying the geographic areas with the largest magnitude of resource needs and allocating resources accordingly. The problem with this approach is that it may result in a suboptimal resource allocation across the national and regional healthcare system, and is essentially reactive in nature. We propose a "value-based" approach, which takes a more proactive and system-wide perspective, based on the long-term value of allocation decisions. Similar perspectives can be seen in the supply chain practice of "demand management" (Croxton et al., 2002), which involves linking and synchronizing demand forecasts with production, procurement, and distribution capabilities to find initiatives which add economic value to the firm's financial performance. Here, instead of measuring economic value added to a firm, we define value in terms of the expected usage of a marginal bed – in other words we seek to obtain, for each new bed added, the greatest use benefit to the greatest number of patients while using the fewest resources possible (Laventhal et al., 2020).

The problem of allocating scarce resources is well-suited to be informed by mathematical optimization models. Mehrotra et al. (2020) developed a stochastic optimization model for the allocation and sharing of ventilators. Billingham et al. (2020) approached the problem of scarce ventilator distribution through a network optimization model. Santini (2020) used integer programming to optimize the allocation of swabs and chemical reagents for COVID-19 testing. However, in time-critical emergency situations, taking an action-oriented perspective is preferred where streamlined, yet informative, "scratch" models are developed to generate insights within a short timeframe (e.g., Kaplan et al., 2020; Manca et al., 2020).

In this spirit, we propose a simple, value-based optimization model for prioritizing hot spots for the allocation of healthcare resources (i.e., new beds) based on an understanding of bed demand and the value of adding additional marginal beds. In this chapter, we describe a simple nonlinear programming model for the decision problem of allocating some number of new hospital beds over time and across a number of geographical locations. The modeling approach is to link the optimization model to epidemiological models which provide uncertain demand forecasts.

## 2. An Optimization Model for Hospital Bed Allocation

The decision problem is to build a limited number of beds across a set of possible locations (modeled at the state level) over multiple time steps (weekly). The number of beds which can be built in a week is limited to a certain maximum capacity. Additionally, it takes some amount of time between when the decision is made to build beds at a location and when they are completed and available for treating new patients. Finally, we assume that the objective of the decision maker in allocating beds is to minimize the total expected shortfall of beds across a given region.

Framed mathematically as an optimization problem, we let the decision variables be denoted by $x(i,t)$, the number of beds to add to state $i$ in week $t$ (therefore there will by $i*t$ decision variables). Let $B_A(i,t)$ be the Beds Available for state $i$ in week $t$. Allowing for a lag time of $n$ weeks, $B_A(i,t+n) = B_A(i,t) + x(i,t)$ (e.g., if we let $n = 3$ weeks, the $x(i,t)$ beds planned to be built in week 1 will be available for use in week 4). The maximum capacity of beds built per week across a region is defined as $x_{cap}(t)$, which functions as a constraint.

Further, let $B_M(i,t)$, $B_L(i,t)$, $B_U(i,t)$ be forecasted Mean Beds Needed, Lower Uncertainty Bound for Beds Needed, and Upper Uncertainty Bound for Beds Needed, respectively, for state $i$ in week $t$, as provided by epidemiological forecasts. The Bed Shortfall then is the difference between Beds Needed and Beds Available, i.e.,

$$B_{SM}(i,t) = B_M(i,t) - B_A(i,t) \qquad \text{Shortfall Mean Beds} \qquad (1)$$

$$B_{SL}(i,t) = B_L(i,t) - B_A(i,t) \qquad \text{Shortfall Lower Beds} \qquad (2)$$

$$B_{SU}(i,t) = B_U(i,t) - B_A(i,t) \qquad \text{Shortfall Upper Beds} \qquad (3)$$

Note that a negative shortfall implies that there is a surplus of beds.

Given a mean, lower, and upper estimate for bed shortfalls, the Expected Shortfall is the probability-weighted average of Shortfall Mean Beds, Shortfall Lower Beds, and Shortfall Upper Beds, based on a selected discrete probability distribution. The weights for Mean, Lower, and Upper were assigned arbitrarily as 0.5, 0.25, and 0.25, respectively:

$$B_s(i,t) = \big(0.5 * \max(0, B_{SM}(i,t))\big) + \big(0.25 * \max(0, B_{SL}(i,t))\big) + \big(0.25 * \max(0, B_{SU}(i,t))\big) \qquad (4)$$

The *max* function is used to ensure that only shortfalls (i.e., situations where demand is greater than capacity) are counted, and ignores situations of surplus beds.

Finally, $B_{s\_total}$ is the Total Expected Shortfall over all states and all weeks, which is the value we seek to minimize, and is defined as:

$$B_{s\_total} = \sum_i \sum_t B_s(i,t) \qquad (5)$$

Thus, the optimization model can be formulated as follows:

| | | | |
|---|---|---|---|
| $MIN$ | $B_{s\_total}$ | Minimize total expected shortfall of beds | (6) |
| $S.T.$ | $\sum_i x(i,t) \leq x_{cap}(t)$ | Capacity constraint | (7) |
| | $x(i,t) \geq 0$ | Non-negativity constraint | (8) |

The objective function (6) states that we seek to minimize the Total Expected Shortfall of beds over all states and all weeks. The first constraint (7) states that the number of beds added for all states $i$ in a region cannot be more than the total regional build capacity for that time period (e.g., per week). Finally, the second constraint (8) states that, naturally, we cannot build a negative amount of beds (or stated differently, we would never decide to remove any beds from any locations).

Finally, following a marginal analysis approach (McKenzie, 1999), The Expected Usage of Marginal Bed is calculated as:

$$\big(0.5 * IF(B_{SM}(i,t) > 0,1,0)\big) + \big(0.25 * IF(B_{SL}(i,t) > 0,1,0)\big) + \big(0.25 * IF(B_{SU}(i,t) > 0,1,0)\big) \quad (9)$$

In other words, these values represent a probability weighted average of whether a shortfall will occur based on mean, lower, and upper estimates of beds needed. Values may equal {1, 0.75, 0.25, 0} and the interpretations are that a marginal bed {will be used, likely to be used, might be used, won't be used}, respectively. Note that the weights of 50%, 25%, and 25% for mean, lower bound, and upper bound, respectively in (9) are used as an approximate discretization of the distribution provided, and are not meant to reflect that the data reflect the 25th, 50th, and 75th percentiles. As a short illustrative example, if demand for beds in a given location on a given date is thought to have a 25% chance of being 1,000 (lower bound), a 50% chance of being 1,500 (mean), and a 25% chance of being 2,000 (upper bound), and current capacity is only 500 beds, then the next marginal bed (i.e., an added 501st bed) will be used on that date with 100% probability, as will the 800th bed and the 999th bed, while the 1,000th bed would have a 75% chance of being used as will the all the beds up to the 1,499th, the 1,500th through 1,999th beds have a 25% chance of being used, and the 2,000th bed and above has a 0% chance of being used on that day.

This is the single time period analysis, but in allocating scarce resources over time, the concern is with the expected number of days a bed will be used. So if the demand peaks and then drops so that over a 3 week period the 1,200th bed has a 100% chance of being used for 1 week, a 75% chance of being used for 2 weeks, and a 25% chance of being used for all 3 weeks, the expectation of the number of weeks the bed would be utilized is (100% * 1) + (75% * 2) + (25% * 3) = 3.25, while the 800th bed would have a similar calculation leading to higher expected use. The expected cumulative use of *n-th* marginal bed in this region would be compared (implicitly, in the course of optimization) against the expected use of beds at different levels in other locations competing for the scarce resources.

## 3. Case Study

### 3.1. Data and Assumptions

Forecast were taken from IHME's "COVID-19 Projections" model (IHME COVID-19 Health Service Utilization Forecasting Team, 2020). Forecasts were obtained from data available as of 25 March, 2020. The data utilized in the optimization model is a subset of the available data, and includes mean hospital beds needed by day, lower uncertainty bound of hospital beds needed by day, upper uncertainty bound of hospital beds needed by day. These data were selected for a subset of north-east U.S. states: Connecticut, Delaware, District of Columbia, Maine, Maryland, Massachusetts, New Hampshire, New Jersey, New York, Pennsylvania, Rhode Island, Vermont, and Virginia. The IHME data were provided daily, but we selected a subset of every 7 days (i.e., weekly) over a 13-week time horizon.

Data on Bed Capacity (i.e., how many beds are available) per state are not included in the downloadable file, but can be found on the IHME website[1], and were entered manually into the spreadsheet (Table 1).

Additional model assumptions and user-defined inputs include the following:

- We assumed that the rate at which new beds can be built per week within the selected region cannot exceed 1,200 beds/week. This rate was estimated based on data reported by the U.S. Army Corps of Engineers[2], where the average construction rate in the study region was approximately 216 beds/week per project, with multiple projects ongoing on any particular day (generally between 4 and 9). Taking six projects per week as a rough mid-point, we arrived at the estimate for 1,200 beds/week.

---

[1] https://covid19.healthdata.org/
[2] https://www.usace.army.mil/Coronavirus/Listing-of-USACE-Contracts-Awarded-for-Alternate-Care-Sites-in-Support-of-COVID-19/

- We assumed the time lag between the decision to build a bed and when it is completed is 2 weeks.

### 3.2. Implementation

The optimization model was implemented in a Microsoft Excel spreadsheet. Since the number of decision variables is large (especially if one chooses to analyze all U.S. states), the default Solver package in Excel is unable to perform the necessary calculations. Given size limitations, the plug-in OpenSolver Advanced[3] was used which can handle large sets of decision variables and allows users to select from a variety of linear and nonlinear solution engines (Mason, 2012). While the problem as formulated above is nonlinear, we implemented the spreadsheet in such a way, through the introduction of extra "dummy" decision variables, that the program could be solved linearly. The COIN-OR CBC (Linear Solver) engine, included in OpenSolver Advanced, was used.

### 3.3. Optimization Results

The decision variables resulting from the optimization analysis are reported for the 25 March data (Table 2). Table 2 shows that in the first week (of 25 March), 748 beds are allocated to NJ, whereas 452 are allocated to NY. In the following weeks, beds are allocated to NJ, MA, DE, ME, MD, NH, and VA, with no beds allocated to any states during or after the week of 13 May.

Figures 1 and 2 show the marginal value of adding an additional bed based on the data set. Figure 1 shows the marginal values before the optimization, i.e., with no beds added anywhere, and Figure 2 shows the marginal values post-optimization. In Figure 1, we see that in the first week, a marginal bed will likely be used in NY and VT. Comparison of the pre- and post-optimization values show that adding beds in the specified locations reduces expected shortfalls and therefore the expected use of a marginal bed, especially later in the time horizon (starting around 06 May). The bottom sections of Figures 1 and 2 shows the remaining expected value of a marginal bed, which is simply the sum of the expected value of a marginal bed from the current and future weeks. Figure 3 plots the cumulative expected use of a marginal bed for each state over time (i.e., the bottom portion of Figure 1). We see that in week 1, New York, Vermont, and New Jersey showed the largest values for the cumulative expected use of a marginal bed. However if we consider the 2 week lag time, and look at week 3, New Jersey, Massachusetts, and New York have the highest values, and in subsequent weeks are surpassed by Delaware, Maine, Maryland and Virginia. From Table 2, we see that beds are allocated to New York and New Jersey early, but then the bulk of the beds are allocated Massachusetts, Maryland, and later to Virginia.

---

[3] https://opensolver.org/

## 4. Discussion

Given the urgent need for insights into the problem of where to allocate new beds and other resources, many simplifying assumptions were made following the "scratch modeling" orientation described above (Kaplan et al., 2020). The fact that many simplifying assumptions were made provides ample opportunities for extensions and future research, which are described in this section.

First, a natural extension of the model described here is the consideration other resources besides hospital beds, including ventilators, intensive care beds, and other PPE. While some patients only need beds, others must be placed on ventilators, and therefore shortfalls of both beds and ventilators should be minimized. The IHME epidemiological model used as a data source provides demand forecasts of hospital beds, ICU beds, and ventilators.

An important consideration is maintaining appropriate staffing levels, including doctors, nurses, and support staff. Even with enough beds and ventilators, a shortage in health care personnel can adversely affect patient outcomes. The U.S. Centers for Disease Control and Prevention (2020) outlined strategies for mitigating healthcare personnel shortages. Maintaining appropriate staffing ratios across regions could be added as a model constraint.

Another area for investigation is in better understanding how beds are used, and therefore how many will be needed. For example, given the cleaning requirements following a patient discharge or death, there may be a percentage of beds which are unused at any point in time. Additionally, as doctors continue to learn how to treat hospitalized patients more effectively and efficiently, bed needs will change. Moreover, as mentioned above, we assumed that beds would never be removed from a location. However, it is possible that beds or ventilators could be relocated. All of these are factors which could be investigated within the existing modeling framework.

A further extension is to link the analysis with geographical modeling tools such as GIS. This would allow analysts to investigate how close facilities are to other hospitals. A natural extension of the model is to consider different geographical scales (e.g., county-level) or metropolitan areas.

Finally, several of the data inputs can be further refined, such as the rate at which beds can be built per week. Subject matter experts will need to be consulted to provide these inputs. These rates might be different across locations/regions depending on a number of factors. In the absence of well-documented data, ranges can be provided and sensitivity analysis performed. Another input consideration is the 0.25-0.5-0.25 probability distribution for expected bed shortfall. Other distributions and parameterizations could be used, such as 0.3-0.4-0.3 (Hurst et al., 2000). Finally, non-linear loss functions may also be investigated.

## 5. Conclusions

In this chapter, we have proposed an approach for identifying "hot spots" where epidemiological models predict the greatest shortfalls in hospital bed capacity. Using an optimization model, we demonstrated the ability to schedule and allocate scarce healthcare resources in specific locations, adding beds such that the total expected shortfall will be minimized. This modeling approach can be used as part of a broader public health strategy, comprising a number of key capabilities. A strategic and proactive allocation of hospital resources, including bed capacity, can aid in an overall effort to increase the resilience of the healthcare system to systemic shocks (Hynes et al., 2020). However, a value-based optimization approach is not limited to the allocation of healthcare resources and can be applied to many different types of disaster response efforts, especially those which continue over a protracted time frame. Disaster response efforts in general could benefit from such resource allocation approaches, especially when used in conjunction with probabilistic forecast models.

While new information about the disease continues to be generated and disseminated by researchers, healthcare practitioners, health agencies, etc., the decision making environment continues to be highly uncertain and dynamic. Public health decisions must be made based on the best current understanding of rapidly evolving datasets and forecasts (Uhlig et al., 2020). However, forecasts can (and do) change, which requires a dynamic and flexible approach to modeling and decision making. With a model like the one described in this chapter, changing forecasts may require iteratively running the model to optimize allocation decisions in the short term, while updating both the forecasts and future allocations regularly as new information becomes available.

**Acknowledgements:** This work was supported by the USACE FLEX Program. The opinions expressed herein are those of the authors alone, and not necessarily of their affiliated institutions.

Table 1: Beds Available by State

| State | Beds Available |
|---|---|
| Connecticut | 1,738 |
| Delaware | 696 |
| District of Columbia | 1,093 |
| Maine | 1,061 |
| Maryland | 3,961 |
| Massachusetts | 4,848 |
| New Hampshire | 1,018 |
| New Jersey | 7,815 |
| New York | 13,010 |
| Pennsylvania | 14,395 |
| Rhode Island | 795 |
| Vermont | 533 |
| Virginia | 6,581 |

Source: https://covid19.healthdata.org/united-states-of-america

Table 2: Optimization Results: Beds Added Based on 25 March Forecast

| State | 3/25/20 20 | 4/1/202 0 | 4/8/202 0 | 4/15/20 20 | 4/22/20 20 | 4/29/20 20 | 5/6/202 0 | 5/13/20 20 | 5/20/20 20 | 5/27/20 20 | 6/3/202 0 | 6/10/20 20 | 6/17/20 20 |
|---|---|---|---|---|---|---|---|---|---|---|---|---|---|
| Connecticut | 0 | 0 | 0 | 0 | 0 | 0 | 0 | 0 | 0 | 0 | 0 | 0 | 0 |
| Delaware | 0 | 0 | 24 | 0 | 0 | 0 | 0 | 0 | 0 | 0 | 0 | 0 | 0 |
| District of Columbia | 0 | 0 | 0 | 0 | 0 | 0 | 0 | 0 | 0 | 0 | 0 | 0 | 0 |
| Maine | 0 | 0 | 12 | 0 | 0 | 0 | 0 | 0 | 0 | 0 | 0 | 0 | 0 |
| Maryland | 0 | 0 | 235 | 1079 | 0 | 0 | 0 | 0 | 0 | 0 | 0 | 0 | 0 |
| Massachuse tts | 0 | 211 | 921 | 0 | 0 | 0 | 0 | 0 | 0 | 0 | 0 | 0 | 0 |
| New Hampshire | 0 | 0 | 8 | 0 | 0 | 0 | 0 | 0 | 0 | 0 | 0 | 0 | 0 |
| New Jersey | 748 | 989 | 0 | 0 | 0 | 0 | 0 | 0 | 0 | 0 | 0 | 0 | 0 |
| New York | 452 | 0 | 0 | 0 | 0 | 0 | 0 | 0 | 0 | 0 | 0 | 0 | 0 |
| Pennsylvani a | 0 | 0 | 0 | 0 | 0 | 0 | 0 | 0 | 0 | 0 | 0 | 0 | 0 |
| Rhode Island | 0 | 0 | 0 | 0 | 0 | 0 | 0 | 0 | 0 | 0 | 0 | 0 | 0 |
| Vermont | 0 | 0 | 0 | 0 | 0 | 0 | 0 | 0 | 0 | 0 | 0 | 0 | 0 |
| Virginia | 0 | 0 | 0 | 121 | 1200 | 1200 | 1200 | 0 | 0 | 0 | 0 | 0 | 0 |

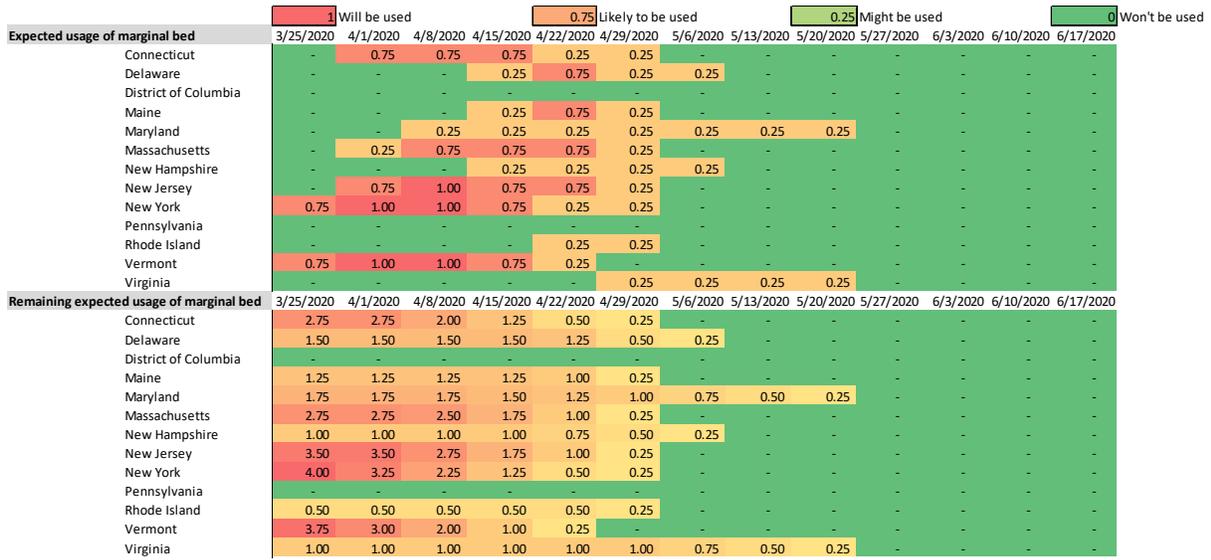

Figure 1: Marginal analysis of beds based on 25 March data (before optimization)

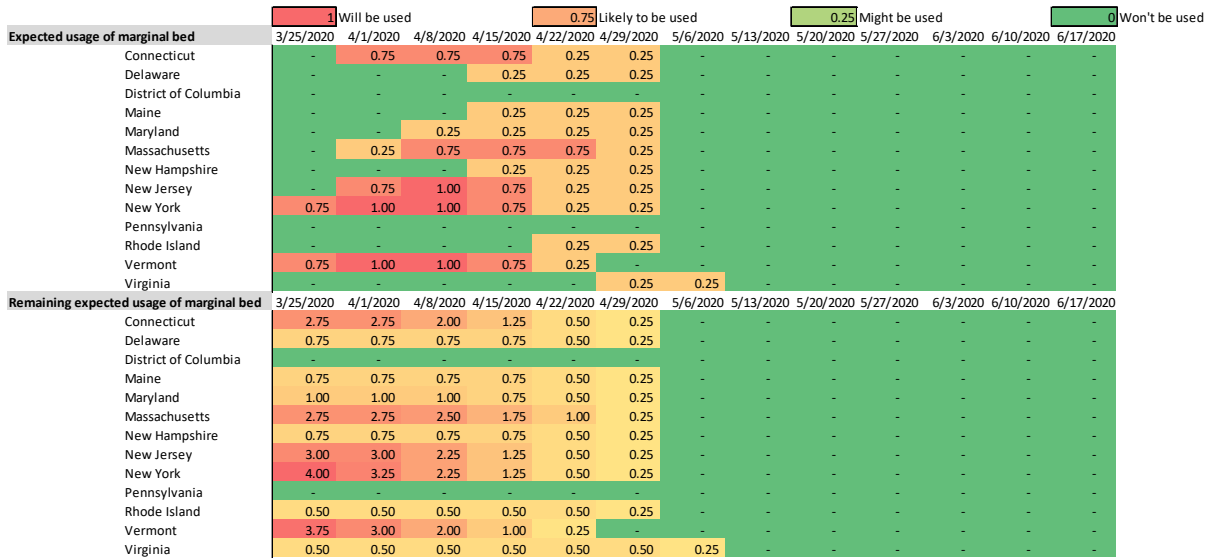

Figure 2: Marginal analysis of beds based on 25 March data (after optimization)

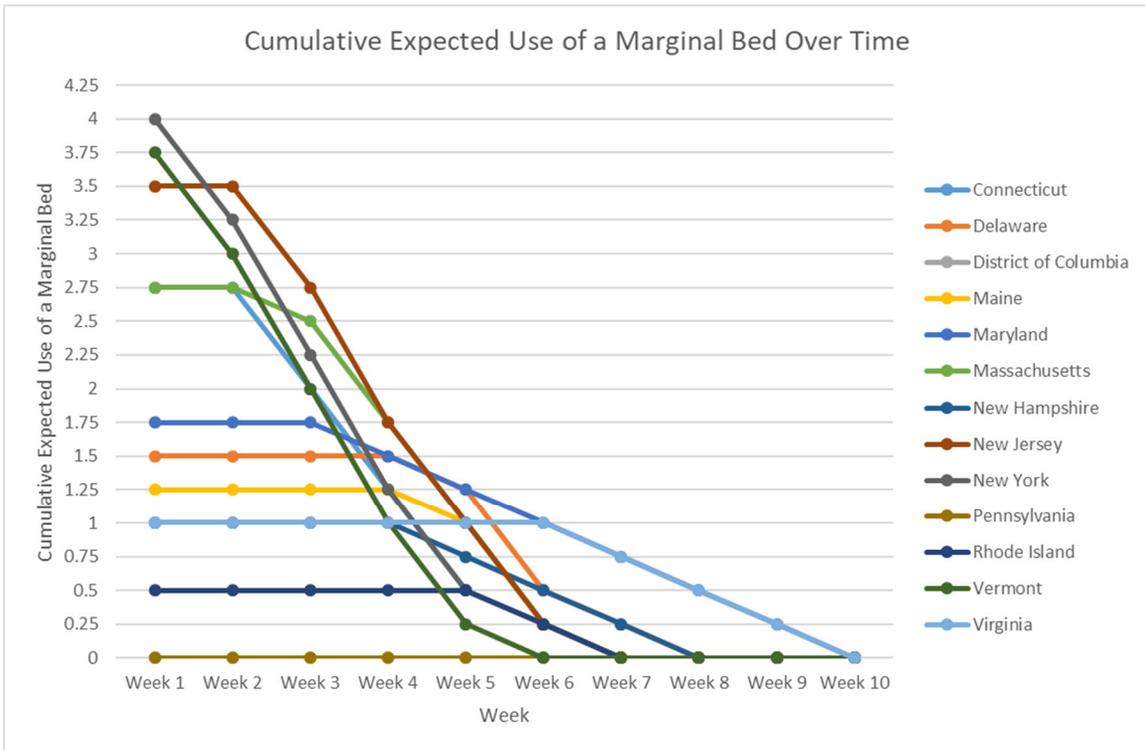

Figure 3: Cumulative expected use of a marginal bed, before optimization